\documentclass{ifacconf}

{}
{}
\newtheorem{remark}{Remark}{}
\newtheorem{assumption}{Assumption}

\usepackage{graphicx}      
\usepackage{natbib}       
\usepackage{graphicx}
\usepackage{graphicx}      
\usepackage{color}
\usepackage{epstopdf} 
\usepackage{graphicx}
\usepackage{pmat}
\usepackage{amsfonts,amsmath,amssymb}
\usepackage{multirow}
\usepackage{mathtools}
\usepackage[most]{tcolorbox}
\usepackage{xcolor}
\usepackage{adjustbox}
\usepackage{epstopdf}
\usepackage{amsmath}
\usepackage{float}
\usepackage[hyphens]{url}
\usepackage{adjustbox}

\newcommand{\Real}{\mathbb R}

\newcommand{\norm}[1]{\left\Vert#1\right\Vert}
\newcommand{\SA}[1]{{\color{black}#1}}

\begin{document}\sloppy
\begin{frontmatter}

\title{Affine Linear Parameter-Varying Embedding of Nonlinear Models with Improved Accuracy and Minimal Overbounding}


\author[First]{Arash Sadeghzadeh} 
\author[Second]{Bardia Sharif}
\author[First,Third]{Roland T\'{o}th} 

\address[First]{Control Systems Group, Department of Electrical Engineering, Eindhoven University of Technology, Eindhoven, The Netherlands(e-mail: a.sadeghzadeh@tue.nl, r.toth@tue.nl).}

\address[Second]{Control Systems Technology Group, Department of Mechanical Engineering, Eindhoven University of Technology, Eindhoven, The Netherlands}

\address[Third]{Systems and Control Laboratory, Institute for Computer Science and Control, Budapest, Hungary}

\begin{abstract}
	In this paper, automated generation of linear parameter-varying (LPV) state-space models to embed the dynamical behavior of nonlinear systems is considered, focusing on the trade-off between scheduling complexity and model accuracy and on the minimization of the conservativeness of the resulting embedding. The LPV state-space model is synthesized with affine scheduling dependency, while the scheduling variables themselves are nonlinear functions of the state and input variables of the original system. The method allows to generate complete or approximative embedding of the nonlinear system model and also it can be used to minimize complexity of existing LPV embeddings. The capabilities of the method are demonstrated on simulation examples and also in an empirical case study where the first-principle motion model of a 3-DOF control moment gyroscope is converted by the proposed method to LPV model with  low scheduling complexity. Using the resulting model, a gain-scheduled controller is designed and applied on the gyroscope, demonstrating the efficiency of the developed approach.  
\end{abstract}
\end{frontmatter}

\section{Introduction}

The \emph{linear parameter-varying} (LPV) framework has been introduced to tackle the control problem of \emph{nonlinear} (NL) and \emph{time-varying} (TV) systems using the extension of powerful linear control methods \cite{RugSha00}.
 This methodology offers great potential in a wide variety of practical applications  \cite{HofWer15}.  By using so-called \emph{scheduling variables}, which express nonlinear or time-varying behavior, LPV systems are capable of representing the solution set of NL/TV systems in terms of a linear structure. Extension of \emph{linear-time invariant} (LTI) methods to exploit this linear proxy representation has seen a tremendous development \cite{HofWer15}; however, the lack of systematic methods that are capable of automatically deriving LPV embeddings of NL/TV systems has prevented the widespread industrial use of the LPV concept \cite{Tot10}. 
 
While LPV system identification methods have matured over the last 15 years with many competitive approaches, e.g. \cite{TotLauGil12,GooPin16,LauTot12,Tot12a,ZhaHuaSu12,BacTotLud14,Tot10,LiuYanZhu19}  to mention a few, conversion methods of existing NL/TV models of applications has seen only moderate progress. As in practice, often high-fidelity models of the target application are available due to the development and design process, e.g. rigid body/flexible motion dynamics in mechatronic applications.  The need for embedding approaches in which such models are converted to LPV description to be used for control design or prediction purposes  is of great significance. 

Existing methods can be categorized as local and global methods. In local LPV model conversion, an NL description of the system is linearized at several operating points, then the obtained linearized models are interpolated to get an LPV model. However, due to the local information on the dynamic aspects, closed-loop stability and performance cannot be guaranteed by controllers, designed based on the resulting LPV models, unless the variation of the scheduling variable is guaranteed to be sufficiently slow, see  \cite{Tot10,BacTotLud14} for an overview. In global methods, the NL/TV system model is directly converted to an LPV representation such that the original system behavior is embedded in the solution set of the resulting LPV model. The methods can be categorized as \emph{substation-based transformation methods} and (automated) \emph{model transformation methods}. In substitution based transformation techniques, NL terms are considered to be absorbed by the introduced scheduling variables, which results in a global LPV embedding of NL dynamics; nevertheless, the applicability of these approaches is either limited to a narrow class of NL systems (e.g., \cite{Shin2002,MarBal04}) or the methods  are based on commonly used, but ad-hoc substitutions (e.g., \cite{Zin2006,Rugh00}). In the so-called velocity linearization \cite{LeiLei98}, differentiating the NL state-space model of a system, a representation in terms of derivatives of the inputs, outputs, and state variables multiplied by some nonlinear functions are obtained.  Considering these nonlinearities as the scheduling, an LPV model is developed, but requiring specialized control synthesis methods. Model transformation methods are based on the systematic exploration of possible ways of reformulating the NL system as an LPV model with the smallest possible conservatism. Next to the computationally intensive methods in \cite{Tot10},  recent developments include approaches based on \emph{linear fractional representation} (LFR) with a nonlinear feedback block  converted to an LPV model depending affinly on the scheduling variables in \cite{SchTot18}. Choosing a LPV embedding for nonlinear systems is investigated in \cite{RobSalBer19} by minimizing the projection of the nonlinearities onto directions deleterious for performance. The later problem is cast as a computationally intensive \emph{linear matrix inequality} (LMI) based optimization. A systematic embedding method to achieve a state-minimal LPV representation in the observable canonical form is presented in \cite{AbbTotPet14}. Besides the problem of LPV embedding, reduction of complexity in sense of reducing the number of scheduling variables, simplification of the dependency on the scheduling variables, and tightening the admissible region of the scheduling variables have received attention in recent years.  In \cite{KwiWer08}, taking advantage of principal component analysis (PCA) applied on the typical scheduling trajectories, a method is  proposed to obtain LPV models with fewer scheduling variables. It is alleged that the procedure can lead to less overbounding of admissible regions for the scheduling variables without providing a rigorous proof. As an extension to that method, a \emph{linear fractional transformation} (LFT)-based LPV representation for descriptor systems is proposed in \cite{HofWer15Conf}. The drawback related to these approaches is that they mainly focus on the scheduling variables not their effects on the dynamical behavior of the system. An approach based on Ho-Kalman algorithm is presented in \cite{SirTot12} to address the problem of joint state-order and scheduling-dependency reduction but applicability of this approach is limited to small scale problems.

In summary, next to accurate representation of the behavior of the NL/TV system, the three main challenges that LPV embedding methods need to face with are (i) \emph{scheduling complexity} minimization, (ii) minimization of \emph{conservativeness} of the embedding, (iii) preservation of \emph{structural properties} like controllability, stability, etc. Most LPV controller synthesis methods rely on linear or quadratic optimization with LMI constraints \cite{Scherer2001,LPVtools2015,CaiCamOli08,DaaBerGer08,SatPea13,SadIET19,Toth19aAUT}, 
the number of which grows exponentially with the scheduling dimension/number of vertices used to describe the scheduling range. Therefore, in terms of problem (i), achieving a minimum number of scheduling variables in LPV modeling has paramount importance in practice. Moreover, many methods are formulated for LPV state-space representations with \emph{affine dependence} on the scheduling variables, while for mechatronic and chemical systems, straightforward manual conversion results in rational or even exponential dependence that is often hidden in new scheduling variables, leading to at least doubling of the scheduling dimension. Regarding problem (ii), the scheduling variables are usually assumed to vary independently in some specified ranges; while as the above discussion exemplify it,  they are functions of some measurable variables of the system with often complicated nonlinear dependence. Thus, the scheduling variable dependency in practice leads to conservativeness of LPV models due to the fact that represented solutions include the solution set of the embedded nonlinear system plus those trajectories that result due to forgetting the above mentioned dependence. This is the price to be paid for a linear representation of the dynamics, but excessive conservativeness can lead to degradation of the achievable performance or even feasibility by LPV control as all extra dynamics resulting purely from conversion are needed to be stabilized and shaped during control synthesis. To the best of our knowledge, a general approach for LPV embedding of nonlinear systems in which the aforementioned factors (i)-(iii) are all appropriately addressed is not available yet. This paper aims to address problems (i)-(ii) and balance complexity, conservativeness and accuracy in LPV model conversion in terms of a practically applicable method.

Three typical problem settings connected to model conversion are considered, namely (a) embedding of nonlinear systems to obtain LPV state-space representation with affine dependence on the scheduling variables and minimal conservativeness; (b) simplifying the dependency of an LPV model depending nonlinearly on the scheduling variables to an affine LPV state-space representation with minimal conservativeness; (c) constructing an affine LPV model with restricted number of scheduling variables from an affine LPV model having too many scheduling variables. Our first contribution is to show that these problem formulations can be uniformly expressed as a single realization problem. Then, by deriving an extension of the approach in \cite{KwiWer08}, we apply \emph{principle component analysis} (PCA) to solve the minimal scheduling variable realization problem under affine dependency of the resulting LPV state-space form using a bundle of generated state and input trajectories of the system along which the variation of state-equations are expressed.  Contrary to the method of \cite{KwiWer08} in which the PCA is applied on the data matrix consisting of the individual scheduling variable trajectories, our contribution is to consider variation of the state-equations directly which, as shown through examples, leads to further reduction of the scheduling complexity. Additionally, an accuracy index is defined to address the trade-off between the number of scheduling variables and the model accuracy, which facilitates determining the number of scheduling variables that are required for the embedding. To minimize conservativeness, we optimize the scheduling range in terms of a minimal hyper-rectangle. Our contribution is to formulate this scheduling range minimization problem and the connected rotational problem of the scheduling space. To reveal the advantages of the proposed method over the existing approaches, the presented method in this paper is applied on simulation examples and validated empirically on a 3-DOF gyroscope system. In the later case, using the obtained LPV model, a full-order gain-scheduled output feedback controller is designed on the converted low complexity model and verified on the experimental setup.

\noindent
\textbf{Notation:} $L_{i,j}\in \Real$ denotes the elements of matrix $L\in \Real^{m\times n}$, i.e.~$L\coloneqq \left[L_{i,j}\right]_{m\times n} $. $\overrightarrow{\Gamma}$ refers to row-wise vectorization of $L$:
\[ \overrightarrow{L}\coloneqq  \left[\begin{array}{cccc} L_{1,1}  \cdots  L_{1,n} & L_{2,1} \cdots L_{2,n}& \cdots  & L_{m,1} \cdots L_{m,n}\end{array}\right]^\top. \] For $X=\overrightarrow{L} \in \Real^{(mn)\times 1}$, the reverse operation is 
\[
L= \underset{\tiny \mbox{$m\times n$}}{\underleftarrow{~~X~~}} \in \Real^{m\times n}.
 \]
The notation $ \norm{L}_F\coloneqq \sqrt{\sum_{i=1}^m\sum_{j=1}^n |L_{i,j}|^2} $ corresponds to the Frobenius norm.
For a vector function $F:\mathbb{R}^{n_\alpha}\rightarrow \Real^{m} $, 
$\text{SD}_{\mathcal{D}_N}(\Gamma(\alpha))$ denotes the empirical standard deviation of the elements of $\Gamma(\alpha(t))$ over a data set $\mathcal{D}_N\coloneqq \{\alpha(t)\}_{t=0}^{N-1}$:
\begin{align*}
&\text{SD}_{\mathcal{D}_N}(\Gamma(\alpha)) \coloneqq  \left[ \begin{array}{cccc} \sigma_{\mathcal{D}_N}(\Gamma_1(\alpha)) &  \cdots & \sigma_{\mathcal{D}_N}(\Gamma_m(\alpha))  \end{array} \right]^\top
\end{align*}
where
\[
\sigma_{\mathcal{D}_N}(\Gamma_i(\alpha))\coloneqq \sqrt{\frac{1}{N}\sum_{t=0}^{N-1}\bigl( \Gamma_i(\alpha(t))- \mathbb{E}_{\mathcal{D}_N}(\Gamma(\alpha))\bigr)^2}
\]
with
$
\mathbb{E}_{\mathcal{D}_N}(\Gamma(\alpha))\coloneqq\frac{1}{N}\sum_{t=0}^{N-1}\Gamma_i(\alpha(t)).
$
 A function $\alpha:\Real \rightarrow \Real$ is called class $\mathcal{C}_1$ \SA{if it is continuous and its first derivative exits}. 

\section{Problem statement}
Consider an NL system defined by the finite dimensional \emph{state-space} (SS) representation:
\begin{subequations} \label{NL-Equ}
	\begin{align}
	\dot{x}(t)&=f(x(t),u(t)), \\
	y(t)&=h(x(t),u(t)),
	\end{align}
\end{subequations}
where $x:\Real \rightarrow \mathbb{X} \subseteq \Real^{n_x}$ is the state variable, $u:\Real \rightarrow \mathbb{U} \subseteq \Real^{n_u}$ is the input , and $y:\Real \rightarrow \mathbb{Y} \subseteq \Real^{n_y}$ is the output of the system for which it is true that $(y,x,u)$ satisfies (\ref{NL-Equ}) in the ordinary sense. $\mathbb{X}$ and $\mathbb{U}$ are considered to be open sets containing the origin. 

\SA{\begin{assumption}
It is assumed that the nonlinear functions $f$ and $h$ are factorisable as 
\begin{subequations} \label{eq:fact}\begin{align}
f(x(t),u(t))&=\mathcal{A}(x(t),u(t))x(t)+\mathcal{B}(x(t),u(t))u(t), \\
h(x(t),u(t))&=\mathcal{C}(x(t),u(t))x(t)+\mathcal{D}(x(t),u(t))u(t),
\end{align}
\end{subequations}
where $\mathcal{A}$, $\mathcal{B}$, $\mathcal{C}$, and $\mathcal{D}$ are bounded and smooth functions on $\mathbb{X} \times \mathbb{U}$. This is a mildly restrictive assumption, as a wide class of nonlinear systems can be represented by  (\ref{eq:fact}), such as rational or polynomial nonlinear systems.
\end{assumption}}
It is supposed that for all initial condition $x_0\in \mathbb{X}$ at any $t_0\in\mathbb{R}$, there exists a unique solution $(y,x,u)$ which is forward complete. Based on these assumptions, denote the solution set,  i.e. the so-called behavior, of the system represented by (\ref{NL-Equ}) as
\begin{multline}
\mathbb{B}_\mathrm{NL}=\{(y,x,u) \in (\mathbb{Y} \times \mathbb{X} \times \mathbb{U})^{\Real^+_0} \mid (y,x,u) \mbox{ s.t.~(\ref{NL-Equ}) holds}  \\ 
   \forall t \in \Real^+_0 \text{ with }  x\in \mathcal{C}_1^{n_x} \text{ and } x(0)=x_0 \in \mathbb{X} \}.
\end{multline} 
Introduce the notation $\mathbb{B}_\mathrm{NL}^{(\mathrm{x})}= \{x\in\mathbb{X}^{\Real^+_0} \mid \exists (y,u) \in (\mathbb{Y} \times  \mathbb{U})^{\Real^+_0}$ $\text{ s.t. } (y,x,u)\in  \mathbb{B}_\mathrm{NL} \}$.
 Factorization \eqref{eq:fact} implies the following SS representation for (\ref{NL-Equ}):
\begin{subequations} \label{StaSpa_rep}
\begin{align}
\dot{x}(t)&=\mathcal{A}(x(t),u(t))x(t)+\mathcal{B}(x(t),u(t))u(t), \\
y(t)&=\mathcal{C}(x(t),u(t))x(t)+\mathcal{D}(x(t),u(t))u(t).
\end{align}
\end{subequations}
In this paper, three types of problems are tackled. 

\textbf{Problem 1} (Embedding a nonlinear model into an affine LPV model): Find an LPV-SS representation for (\ref{StaSpa_rep}) in the form of
\begin{subequations} \label{StaSpa-Prob1}
	\begin{align}
		\dot{x}(t)&= A(\theta(t))x(t)+B(\theta(t))u(t),\\
		y(t)&=C(\theta(t))x(t)+D(\theta(t))u(t),
	\end{align}
\end{subequations} 
such that $\theta\coloneqq\mu(x,u)$ $\mu:\mathbb{X}\times\mathbb{U}\rightarrow\Theta \subseteq \Real^{n_\theta}$ is a bounded smooth vector function, $\Theta$ is a convex set, and matrix functions $A$, $B$, $C$, and $D$ have affine dependence on $\theta(t)\coloneqq[\theta_1(t) \cdots \theta_{n_{\theta}}(t)]^\top$, i.e.,
\begin{equation} \label{Aff_rep}
M(\theta(t))=M_0+\sum_{i=1}^{n_\theta} \theta_i(t) M_i, \\
\end{equation}
where $M(\theta(t))$ represents any of the matrix functions $A(\theta(t)),\ldots,$ $D(\theta(t))$. Let us define the solution set of (\ref{StaSpa-Prob1}) as follows: 
\begin{multline}
\mathbb{B}_\mathrm{LPV}=\{(y,x,u,\theta) \in (\mathbb{Y} \times \mathbb{X} \times \mathbb{U} \times \Theta)^{\Real^+_0} \mid (y,x,u,\theta) \\ \mbox{s.t.~(\ref{StaSpa-Prob1}) }  
  \mbox{holds } \forall t \in \Real^+_0 \text{ with }  x\in \mathcal{C}_1^{n_x} \text{ and } x(0)=x_0 \in \mathbb{X} \}.
\end{multline} 
Note that $\mathbb{B}_\mathrm{NL}\subseteq\mathbb{B}_\mathrm{LPV}$, due to the fact that $\theta$ can get any value in the admissible set $\Theta$ irrespective of the values of $x$ and $u$.  For this problem, the goal is to minimize $\mathbb{B}_{\mathrm{LPV}}\backslash \mathbb{B}_{\mathrm{NL}}$ in terms of a measure on $\mathbb{X}\times \mathbb{U}$ among all possible choices of $A,\ldots,D$, $\mu$, and $\Theta$. 

Additionally, one can alternatively seek an approximate affine LPV model as follows:
\begin{subequations}\label{App_Aff_LPV}
	\begin{align} 
	\dot{x}(t)&=\hat{A}(\hat{\theta}(t))x(t)+\hat{B}(\hat{\theta}(t))u(t), \\
	y(t)&=\hat{C}(\hat{\theta}(t))x(t)+\hat{D}(\hat{\theta}(t))u(t),
	\end{align}
\end{subequations}
where $\hat{\theta} \coloneqq \hat{\mu}(x,u): \mathbb{X}\times \mathbb{U} \rightarrow \hat{\Theta} \subseteq  \Real^{n_{\hat{\theta}}}$ and $n_{\hat{\theta}}<n_\theta$ by minimizing a measure of discrepancy between the matrices $\mathcal{A},\ldots,$ $\mathcal{D}$  and  $\hat{A},\ldots,\hat{D}$, which measure is precisely defined later in this section.

\textbf{Problem 2} (Converting an LPV model  with nonlinear dependency on the scheduling variables into an affine LPV model): Given an LPV embedding of (\ref{NL-Equ}) in the form of 
\begin{subequations} \label{StaSpa-Prob2}
	\begin{align}
	\dot{x}(t)&=\tilde{\mathcal{A}}(\alpha(t))x(t)+\tilde{\mathcal{B}}(\alpha(t))u(t), \\
	y(t)&= \tilde{\mathcal{C}}(\alpha(t))x(t)+\tilde{\mathcal{D}}(\alpha(t))u(t),
	\end{align}
\end{subequations}
for which $\tilde{\mathcal{A}}$, $\tilde{\mathcal{B}}$, $\tilde{\mathcal{C}}$, and $\tilde{\mathcal{D}}$ are non-affine functions of the scheduling variable $\alpha=\kappa(x,u)$ with $\alpha(t) \in \Omega_\alpha\subset \Real ^n_\alpha$.
 In practice, such an embedding can easily happen in case of manual conversion of an NL model to an LPV form (e.g. in case of mechatronic systems (see Sec.\ \ref{sec:exp}), direct substitution of the position dependency in the inertia matrices with scheduling variables leads to LPV-SS representations with rational dependence). The goal is to find $A,\ldots,D$, $\mu$, and $\Theta$ in terms of (\ref{StaSpa-Prob1}) and (\ref{Aff_rep}) such that the discrepancy between the matrices $\tilde{\mathcal{A}},\ldots,\tilde{\mathcal{D}}$ and $A,\ldots,D$ is minimized. As in our setting LPV models represent an underlying NL system, we will consider this minimization  over all  realizations of $\alpha$ and $\theta$ in terms of $\kappa(\mathbb{B}_\mathrm{NL}^{(\mathrm{x,u})})$ and $\mu(\mathbb{B}_\mathrm{NL}^{(\mathrm{x,u})})$.

\textbf{Problem 3} (Obtaining an affine LPV model with reduced number of scheduling variables from  an affine LPV model): Given an LPV embedding of (\ref{NL-Equ}) in terms of (\ref{StaSpa-Prob2}), where the state-space matrices are affine functions of $\alpha$. The goal is to find an approximate  LPV model (\ref{App_Aff_LPV}) such that the discrepancy between the matrices $\tilde{\mathcal{A}},\ldots,\tilde{\mathcal{D}}$ and $\hat{A},\ldots,\hat{D}$ is  minimized over all  realizations of $\alpha$ and $\hat{\theta}$ in terms of $\kappa(\mathbb{B}_\mathrm{NL}^{(\mathrm{x,u})})$ and $\hat{\mu}(\mathbb{B}_\mathrm{NL}^{(\mathrm{x,u})})$. This problem represents the case when the initial number of scheduling variables $n_\alpha$ and/or the admissible set for $\alpha$ are non-minimal or further reduction of these is required for feasibility of control synthesis.

Indeed, the mentioned problems are all low complexity embedding problems for either nonlinear or complex LPV systems. The objective is to find an LPV model, affine with respect to a set of constructed scheduling variables, while the accuracy and the conservativeness of the resulting embedding is taken into account. 
Next we show that Problems 1-3 can be considered under a unified setting. Introduce the representation
\begin{equation} \label{True-Model}
\left[\begin{array}{c} 
\dot{x}(t) \\ y(t)
\end{array} \right]= L(\alpha(t)) 
\left[\begin{array}{c} 
x(t) \\ u(t)
\end{array} \right],
\end{equation}
  where 
\[
L(\alpha(t))\coloneqq \left[L_{i,j}(\alpha(t))\right]_{m\times n} \in \Real^{m\times n},
\]
with
$
m\coloneqq n_x+n_y$, $n\coloneqq n_x+n_u.
$
The variables $\alpha_i(t)$ are assumed to lie in a hyper-rectangle $\Omega_\alpha$, which is the Cartesian product of intervals
\[
\underline{\alpha}_i\leq\alpha_i(t)\leq\overline{\alpha}_i
\]
where $\underline{\alpha}_i$ and $\overline{\alpha}_i$ are a priori known. Thus, $\alpha(t) \in \Omega_\alpha,~\forall t\geq0$. 
Then Problems 1-3 are represented as follows:
\begin{itemize}
	\item{Problem 1:} $L(\alpha(t))$ is a NL matrix function of $\alpha(t)\coloneqq [ x(t)^\top~u(t)^\top]^\top$ and is defined as follows:
	\begin{equation}
	L(\alpha(t))\coloneqq\left[ \begin{array} {cc}
	\mathcal{A}(x(t),u(t)) & \mathcal{B}(x(t),u(t)) \\
	\mathcal{C}(x(t),u(t)) & \mathcal{D}(x(t),u(t))
	\end{array}
	\right].
	\end{equation}
	\item{Problem 2:} $L(\alpha(t))$ is a NL matrix function of the  scheduling variable $\alpha(t)$ and defined as follows:
	\begin{equation} \label{L_p2}
	L(\alpha(t))\coloneqq \left[ \begin{array} {cc}
	\tilde{\mathcal{A}}(\alpha(t)) & \tilde{\mathcal{B}}(\alpha(t)) \\
	\tilde{\mathcal{B}}(\alpha(t)) & \tilde{\mathcal{C}}(\alpha(t)) 
	\end{array}
	\right].
	\end{equation}
	\item{Problem 3:} $L(\alpha(t))$ is an affine function of  scheduling variable $\alpha(t)$ and is defined as in (\ref{L_p2}).  
\end{itemize}

Let us define
\begin{align*}
\Gamma(\alpha(t))\coloneqq&  \left[\begin{array} {cccc} \Gamma_1(\alpha(t)) & \Gamma_2(\alpha(t)) & \cdots & \Gamma_{n_\Gamma}(\alpha(t))\end{array}\right]^\top \\
=&\overrightarrow{L}(\alpha(t)) \in R^{n_\Gamma},
\end{align*}
where $n_\Gamma=(n_x+n_y)(n_x+n_u)$. 
The ultimate goal in this paper is to find an LPV model with affine dependency 
\begin{equation} \label{App-model}
\left[\begin{array}{c} 
\dot{x}(t) \\ y(t)
\end{array} \right]= \hat{L}(\theta(t)) 
\left[\begin{array}{c} 
x(t) \\ u(t)
\end{array} \right],
\end{equation}
for (\ref{True-Model}) by introducing an affine mapping 
\begin{align} \nonumber \label{T_map}
\theta(t)=&\left[\begin{array} {ccc} \theta_1(t) &  \cdots & \theta_{n_\theta}(t) \end{array} \right]^\top \\
\coloneqq& \mathcal{T} (\Gamma(\alpha(t))) \in \Omega_\theta, \quad \mathcal{T}: \Real^{n_\Gamma}\rightarrow\Real^{n_\theta}
\end{align}
 such that an  accuracy index (defined next) is minimized for a prescribed value of $n_\theta\leq n_\Gamma$. Alternatively, the number of scheduling variables $n_\theta$ can be chosen by the designer based on the accuracy index value obtained for different number of scheduling variables. The hyper-rectangle set  $\Omega_\theta$ denoted by
\begin{equation} \label{LU_theta}
\underline{\theta}_i\leq \theta_i (t) \leq \overline{\theta}_i,
\end{equation}
is characterized by its  lower  and upper bounds $\underline{\theta}_i$ and  $\overline{\theta}_i$. The mapping  $\mathcal{T} (\Gamma(\alpha(t)))$ should be determined in such a way that  the volume of $\Omega_\theta$  is kept at its possible minimum to minimize the conservativeness of the obtained model.

Under the assumption that a $\mathcal{D}_N\coloneqq \{\alpha(t)\}_{t=0}^{N-1}$ data set, representative w.r.t. typical operation of the system \eqref{True-Model} (bundle of measured trajectories, equidistant griding of $\mathbb{X}\times\mathbb{U}$, etc.)  is available, we chose the accuracy index as the weighted norm:
\begin{equation} \label{Acc_Crit}
\eta \coloneqq\norm{W(\Pi_\alpha-\hat \Pi_\theta)}_F 
\end{equation}
where  
\begin{subequations}
\begin{align}
\Pi_\alpha&\coloneqq \left[ \begin{array}{cccc} \label{PAIalfa}
\Gamma(\alpha(0)) & \Gamma(\alpha(T))   & \cdots &\Gamma(\alpha((N-1)T)) \\
\end{array}\right], \\
\hat\Pi_\theta& \coloneqq \left[ \begin{array}{cccc}
\hat{\Gamma}(\theta(0)) &\hat{ \Gamma}(\theta(T))   & \cdots &\hat{\Gamma}(\theta((N-1)T)) \\
\end{array}\right],
\end{align}
\end{subequations}
with 
$
\hat{\Gamma}(\theta(t))=\overrightarrow{\hat{L}}(\theta(t)) \in R^{n_\Gamma}
 $, sampling time $T>0$ and
weighting
\begin{equation} \label{Weight_fun}
W\coloneqq\text{diag}\left( \text{SD}_{\mathcal{D}_N}\left(\Gamma(\alpha\right)\right)^{-1}.
\end{equation}
\SA{Note that the Frobenius norm of a matrix is a convenient metric to quantify the approximation error of matrices based on  \emph{singular value decomposition} (SVD) type of projections and makes it possible to give explicit characterization of their approximation error. As it becomes clear later, our proposed method is based on such an SVD type of approximation, hence, the Frobenius norm is a natural choice of performance measure for our method.  However, it is just a particular choice of norm and other matrix norms can be utilized for this purpose, although other choices lose the connection with the SVD based projection.} In the subsequent sections, the problem of finding the mapping  $\mathcal{T}$, introduced in (\ref{T_map}),  is addressed.

\section{Parameter set mapping}
To compute the scheduling mapping, PCA is applied on the data matrix (\ref{PAIalfa}), inspired by the method in \cite{KwiWer08}. Note that contrary to the method in \cite{KwiWer08}, PCA is applied on the data matrix $\Pi_\alpha$, capturing the time-domain trajectories of all the elements of the state-space matrices, not only the scheduling trajectories. As we will show, this not only allows to jointly treat Problems 1-3, but it also allows  achieving more accurate models with less number of scheduling variables compared to \cite{KwiWer08}. 

First, let us define the affine projection law $\mathcal{N}^{(i)}$ as follows: 
\begin{multline}
\bar{\Gamma}_i(\alpha(t))\coloneqq 
\mathcal{N}^{(i)}(\Gamma_i(\alpha(t)))\\ = \frac{1}{\sigma_{\mathcal{D}_N}(\Gamma_i(\alpha))}\bigl( \Gamma_i(\alpha(t))-\mathbb{E}_{\mathcal{D}_N}(\Gamma(\alpha))\bigr)
\end{multline}
where $\sigma_{\mathcal{D}_N}(\Gamma_i(\alpha))$ and $\mathbb{E}_{\mathcal{D}_N}(\Gamma(\alpha))$ are respectively  the standard deviation and the mean of $\Gamma_i(\alpha)$ over the data set $\mathcal{D}_N$. Now, we can define
$
\bar{\Gamma}(\alpha(t))=\mathcal{N}(\Gamma(\alpha(t)))
$
which means that the corresponding affine mappings are applied on the related elements of the vector $\Gamma(\alpha(t))$. Similarly, we can introduce $\bar{\Pi}_\alpha^{(i)}=\mathcal{N}^{(i)}(\Pi_\alpha^{(i)}) 				
$ and the overall $\bar{\Pi}_\alpha=\mathcal{N}(\Pi_\alpha)$
 to obtain a scaled (unit variance), zero mean representation of the variation of the state-space matrices. 
 Let 
\begin{equation}
\bar{\Pi}_\alpha=U\Sigma V^\top
\end{equation}
be the SVD of $\bar{\Pi}_\alpha$. Singular values indicate the principal components of the data. Small singular values indicate relatively unimportant components \cite{OlvSha06}, which means that projection onto a low-dimensional subspace spanned by the dominant singular vectors will not cause losing too much information. 
 Suppose that $\sigma_1,\sigma_2,\cdots,\sigma_{n_\rho}$ are considered as the significant singular values (based on their relative magnitudes). By neglecting the singular values $\sigma_{n_\rho+1},\cdots,\sigma_{n_\Gamma}$ and  partitioning $U$, $\Sigma \coloneqq \text{diag}(\sigma_1,\sigma_2,\cdots,\sigma_{n_\rho},\cdots,\sigma_{n_\Gamma})$, and V as follows:
\[
\Sigma\coloneqq\left[\begin{array}{ccc} \Sigma_\rho & 0 & 0\\ 0 & \Sigma_\eta & 0 \end{array}\right],~U\coloneqq \left[ \begin{array}{cc} U_\rho & U_\eta \end{array}\right], ~V\coloneqq \left[ \begin{array}{cc} V_\rho & U_\eta \end{array}\right],
\]
where $\Sigma_\rho=\text{diag}(\sigma_1,\sigma_2,\cdots,\sigma_{n_\rho})$ and $\Sigma_\eta=\text{diag}(\sigma_{n_\rho+1},\cdots,$ $\sigma_{n_\Gamma}) $, one can obtain the following approximation for $\Pi_\alpha$
\begin{equation} \label{PiHatAlpha}
\hat\Pi_\rho\coloneqq\mathcal{N}^{-1} \left( U_\rho U_\rho^\top \bar{\Pi}_\alpha\right)=\mathcal{N}^{-1} \left( U_\rho U_\rho^\top \mathcal{N}(\Pi_\alpha)\right) \approx \Pi_\alpha,
\end{equation}
where $\mathcal{N}^{-1}$ denotes the rescaling and translation, respectively, such that $\mathcal{N}^{-1}\left( \mathcal{N}\left( \Pi_\alpha\right)\right)=\Pi_\alpha$.

\SA{Let us define the} affine reduced mapping
\begin{equation} \label{Rho}
\rho(t)\coloneqq \mathcal{M}(\alpha(t))=U_\rho^\top \bar{\Gamma}(\alpha(t)) = U_\rho^\top \mathcal{N}(\Gamma(\alpha(t))),
\end{equation}
\SA{considering (\ref{PAIalfa}) and (\ref{PiHatAlpha}), one can see that }
\[
\Pi_\alpha  \approx  \hat\Pi_\rho\coloneqq  \left[\begin{array}{cccc} \hat{\Gamma}(\rho(0)) & \hat{\Gamma}(\rho(T)) & \cdots & \hat{\Gamma}(\rho((N-1)T))\end{array}\right]
\]
where 
\begin{equation} \label{Gamma_hat_rho}
\hat{\Gamma}(\rho(t))\coloneqq \mathcal{N}^{-1}\left(U_\rho\rho(t)\right).
\end{equation}
Subsequently, one can define
\begin{equation} \label{L_hat_rho}
\hat{L}(\rho(t)) \coloneqq \underset{\tiny \mbox{$m\times n$}}{\underleftarrow{~~\hat{\Gamma}~~}}(\rho(t)).
\end{equation}
Note that $\hat{L}(\rho(t))$ is an affine function of $\rho(t)$, and $\rho(t)$ is also an affine function of $\Gamma(\alpha(t))$. Thus, $\hat{L}(\rho(t))$ is an affine function of $\Gamma(\alpha(t))$, but depending on the function $\Gamma$ (original dependencies of the state-matrices) can be a nonlinear function of $\alpha$.
Based on a well-known matrix approximation lemma \cite{EckYou36}, we have
\begin{equation}
\norm{\mathcal{N}(\Pi_\alpha)-\mathcal{N}(\hat \Pi_\rho)}_F =\sigma_{n_\rho+1}+\cdots+\sigma_{n_\Gamma}\coloneqq \eta
\end{equation}
Bear in mind that $\mathcal{N}$ is an affine transformation. Therefore, one can easily see that
$
\mathcal{N}(\Pi_\alpha)-\mathcal{N}(\Pi_\rho)=W(\Pi_\alpha-\Pi_\rho)
$
where the matrix $W$ is given by (\ref{Weight_fun}). This immediately implies
\begin{equation} \label{Acc_medium}
\eta=\sigma_{n_\rho+1}+\cdots+\sigma_{n_\Gamma}=\norm{W(\Pi_\alpha-\Pi_\rho)}_F.
\end{equation}

Note that $\rho(t)\coloneqq\left[\begin{array} {ccc} \rho_1(t) &  \cdots & \rho_{n_\rho}(t) \end{array} \right]^\top\in \Real^{n_\rho}$ belongs to a hyper-rectangle $\Omega_\rho$ denoted by
\[
\underline{\rho}_i\leq \rho_i(t) \leq \overline{\rho}_i.
\]
The lower and upper bounds $\underline{\rho}_i$ and $\overline{\rho}_i$ are obtained respectively as the minimum and maximum values of  $\rho_i(t)$ in terms of (\ref{Rho}) over all admissible values of $\alpha(t)\in \Omega_\alpha$.   

Taking into account that $\hat{L}(\rho(t))$ depends affinely on the newly introduced scheduling vector $\rho(t)$, which is in turn an affine vector function of the elements of $L(\alpha(t))$, one can see that the embedding Problems 1-3 have been successfully addressed. In the next section, the reduction of the associated conservativeness with the LPV models is investigated. 

\begin{remark}
 Note that the number of elements of $L(\alpha(t))$  in Problem 1 is usually greater than the number of individual nonlinear functions that appear in $L(\alpha(t))$. In this case, some of the singular values become zero. If $n_\rho$ is chosen equal to the number of nonzero singular values, then an exact LPV model is obtained (see \ref{Example2}), and an approximate LPV model is obtained if $n_\rho$ is selected smaller than the number of nonzero singular values. However, for Problems 2 and 3, usually an approximate affine LPV model is developed where the approximation error is characterized by the accuracy index (\ref{Acc_Crit}). 
\end{remark}

\section{Minimal enclosing hyper rectangles}

The hyper rectangle set $\Omega_\rho$ contains the new scheduling variable $\rho(t)$; however, $\Omega_\rho$ is not necessarily the hyper rectangle with the smallest volume. In this section, we would like to introduce an invertible affine transformation $\mathcal{R}$ consisting of translation and rotation as follows:
\[
\theta(t)\coloneqq \mathcal{R} (\rho(t)) \in \Omega_\theta, \quad \mathcal{R}: \Real^{n_\rho}\rightarrow\Real^{n_\theta},\quad n_\rho=n_\theta
\]
such that the hyper rectangle $\Omega_\theta$, denoted by (\ref{LU_theta}), has the minimum possible volume. The problem to find the hyper rectangle $\Omega_\theta$ with the minimum possible volume can be cast as the problem of finding a hyper rectangle with minimum-volume enclosing a set of points. We consider two distinct cases $n_\theta\leq3$ and $n_\theta>3$ for reasons that will be clear soon. 

\subsection{The case when $n_\theta\leq3$}

  Finding the minimum volume enclosing hyper rectangle  has already been tackled for two- and three-dimensional point sets in the literature \cite{FreSha75,Tou83,Rou85}.   
 In  \cite{BarHar01}, finding a minimum-volume bounding box for a set of $n$ points in $\Real^3$ is considered where an efficient $\mathcal{O}(n+1/\epsilon^{4.5})$-time algorithm for computing a $(1+\epsilon)$-approximation of the minimum-volume bounding box  is solved; thus, the running time of this algorithm is linear in $n$ (number of the time samples). Examples of implementation of this algorithm can be found in \cite{Kor20, Die20}. 
   The minimum-volume bounding box obtained by any of the available methods is  characterized by its vertices. Note that the bounding box is generally not aligned with the coordinate axes. To describe it with lower and upper bounds on the individual scheduling variables, which is  more desirable from the viewpoint of controller synthesis, one should resort to an appropriate rotation of the scheduling space. Thus, we first find the minimum volume enclosing hyper rectangle, then, we compute a rotation of the scheduling space, i.e. transformation of  $\rho(t)$, such that we align the hyper rectangle with the new scheduling coordinate axes. 
 
  To compute the rotation transformation, we take advantage of the Kabsch algorithm \cite{Kab76}. Let us define
 \[
 P\coloneqq \left[\begin{array}{cccc}
 v_1-\bar{v} & v_2-\bar{v} & \cdots & v_{n_v}-\bar{v}
 \end{array}\right]
 \]
where $v_i$ is either in $\Real^2$ or $\Real^3$ and represents vertex $i$ of the minimum-volume bounding box with $n_v$ vertices (either $n_v=4$ or $n_v=8$) and 
$
\bar{v}=\frac{1}{n_v}\sum_{i=1}^{n_v} v_i
$
is the centroid of the box. Let us  consider 
\[
Q\coloneqq\left[ \begin{array}{rrrr}
\bar{\sigma}_1  & \bar{\sigma}_1   & -\bar{\sigma}_1 & -\bar{\sigma}_1 \\
\bar{\sigma}_2 & -\bar{\sigma}_2 & -\bar{\sigma}_2 & \bar{\sigma}_2 \end{array} 
\right]
\]
for $n_\theta=n_\rho=2$ scheduling variables and
\[
Q\coloneqq\left[ \begin{array}{rrrrrrrr}
\bar{\sigma}_1   & \bar{\sigma}_1   & -\bar{\sigma}_1 & -\bar{\sigma}_1 & -\bar{\sigma}_1 & -\bar{\sigma}_1 & \bar{\sigma}_1        & \bar{\sigma}_1  \\
\bar{\sigma}_2  & -\bar{\sigma}_2  & -\bar{\sigma}_2 & \bar{\sigma}_2 & \bar{\sigma}_2  & -\bar{\sigma}_2 & -\bar{\sigma}_2      & \bar{\sigma}_2\\
\bar{\sigma}_3  & \bar{\sigma}_3   & \bar{\sigma}_3   & \bar{\sigma}_3 & -\bar{\sigma}_3  & -\bar{\sigma}_3   & -\bar{\sigma}_3   & -\bar{\sigma}_3\end{array} 
\right]
\]
for $n_\theta=n_\rho=3$ scheduling variables, where $\bar{\sigma}_1,\bar{\sigma}_2$ and $\bar{\sigma}_1,\bar{\sigma}_2,\bar{\sigma}_3$ are the singular values of $P$ for the case of $n_\theta=2$ and $n_\theta=3$, respectively. In virtue of the Kabsch algorithm, one can obtain the transformation $\mathcal{R}$ having been previously introduced as follows:
\begin{equation} \label{R1}
\theta(t)= \mathcal{R}(\rho(t))=(QP^\top PQ^\top)^{\frac{1}{2}}(PQ^\top)^{-1} \left( \rho(t)-\bar{v}\right)+\bar{v}
\end{equation}
This way the newly introduced scheduling variable vector $\theta(t)$ reside in a hyper rectangle $\Omega_\theta$ which is defined by (\ref{LU_theta}), i.e. the hyper rectangle $\Omega_\theta$ can be characterized by the lower and upper bounds on the individual scheduling variables $\theta_i$ . It should be mentioned that the volume of $\Omega_\theta$ is exactly equal to the volume of the   initial minimum-volume bounding box due to the fact that $\mathcal{R}(\cdot)$ consists of  rotation and translation only which operations preserve volume. 

\begin{remark}
	It is worth mentioning that the rotated minimum-volume bounding box, called hereafter $\Omega_\theta$, is not unique due to the fact that the rotation may be carried out in different directions by choosing alternative ordering in $Q$. However, irrespective of the direction  of the rotation, the obtained minimum-volume bounding boxes can be characterized by the upper and lower bounds of the introduced scheduling variables. These bounding boxes can be  converted to each other by changing the ordering of the scheduling variables  and/or changing the sign of the variables.  
\end{remark}

\subsection{The case when $n_\theta>3$}
We are not aware of any previously-published polynomial-time algorithm that tackles the problem of finding an enclosing hyper rectangle for a set of points for dimension higher than 3. In case it is required to have an LPV model with more than three scheduling variables, one solution is to find a minimum-volume ellipsoid enclosing the scheduling variable trajectories $\rho(t)$ corresponding to the variation in $\mathcal{D}_N$, and then to determine a  hyper rectangle enclosing this ellipsoid to be considered as the minimum-volume box. However, there is no guarantee that this hyper rectangle will be the minimum-volume hyper rectangle which encloses the possible scheduling  variations, but it may potentially have a volume less than $\Omega_\rho$, resulting in a model with reduced conservativeness.

The problem of finding the minimum-volume enclosing ellipsoid has been widely investigated in the literature, see e.g. \cite{Kha96,KumYil05,TodYil07}. An  algorithm to solve this problem  can be found in \cite{Mos20}.   Suppose that the minimum-volume ellipsoid is parametrized as follows:

\[
\mathcal{E}\coloneqq\left\{ v\in \Real^{n_\theta}\mid  (v-\bar{v}_e)^\top P_e (v-\bar{v}_e)\leq 1\right\}
\]
where $n_\theta > 3$ and $P_e\in \Real^{n_\theta \times n_\theta}$ is a symmetric positive-definite matrix and $\bar{v}_e \in \Real^{n_\theta}$ is the center of the ellipsoid. In short, this ellipsoid can be obtained by solving the following convex optimization problem:
\begin{align}
\min_{P_e,\bar{v}_e} & \quad -\log\det P_e \\ \notag
\mbox{s.t.} & \quad 
(v-\bar{v}_e)^\top P_e (v-\bar{v}_e) \leq 1,\quad \forall v\in \hat{ \mathcal{D}}_N \\ \nonumber
& \quad P_e>0
\end{align}
where $\hat{\mathcal{D}}_N\coloneqq \left \{ \rho(t)\right\}_{t=0}^{N-1}$, and $P_e$ and $\bar{v}_e$ are decision variables. Let
\[
P_e=U_e\Sigma_eV_e^\top
\]
be the singular value decomposition of $P_e$. Then, using the following transformation
\begin{equation} \label{R2}
\theta(t)=\mathcal{R}(\rho(t))=U_e(\rho(t)-\bar{v}_e)+\bar{v}_e
\end{equation}
 the minimum-volume ellipsoid is rotated around the center $\bar{v}_e$ such that the principal axes become parallel to the coordinate axes. Now, the required hyper rectangle can be defined by the lower and upper bounds on the individual scheduling variables, which are obtained as the minimum and maximum values of $\theta_i$.

\section{Affine LPV  model construction}
To recapitulate the previous sections, one can obtain an approximate affine LPV model (\ref{App-model})  with $n_\theta$ scheduling variables for any of the three considered problems  by the following affine mapping
\begin{equation*}  
\theta(t) \coloneqq \mathcal{T}(\alpha(t))=\mathcal{R}\left( \mathcal{M}(\alpha(t))\right)=\mathcal{R}\left( 
U_\rho^\top \mathcal{N}\left(\Gamma(\alpha(t))\right) \right)\in \Omega_\theta
\end{equation*}
where $\mathcal{R}(\rho(t))$ is given by either (\ref{R1}) or (\ref{R2}). The hyper rectangle  $\Omega_\theta$  is characterized by the upper and lower bounds of $\theta_i$. 

The accuracy index for the approximate model is as follows:
\begin{equation}  \label{Acc_Final}
\eta =\norm{W(\Pi_\alpha-\Pi_\theta)}_F=\sigma_{n_\rho+1}+\cdots+\sigma_{n_\Gamma}\
\end{equation}
where $\Pi_\theta=\Pi_\rho|_{\rho=\mathcal{R}^{-1}(\theta)}$. Note that $\theta(t)=\mathcal{R}(\rho(t))$ is just a change of variable; consequently, it is easy to see that $\Pi_\theta=\Pi_\rho$. Thus, (\ref{Acc_medium}) immediately implies (\ref{Acc_Final}). As we mentioned previously, $\mathcal{M}(\cdot)$ is an affine mapping; therefore, $\mathcal{T}(\cdot)$ is also an affine mapping.   

Taking into account (\ref{Gamma_hat_rho}), (\ref{L_hat_rho}), (\ref{R1}), and (\ref{R2}) and by defining
\[
\hat{\Gamma}(\theta(t))\coloneqq \mathcal{N}^{-1}\left(U_\rho \mathcal{R}^{-1}\left(\theta(t)\right)\right),
\]
the  affine LPV model (\ref{App-model}) is  obtained where
\[
\hat{L}(\theta(t))\coloneqq \underset{\tiny \mbox{$m\times n$}}{\underleftarrow{~~\hat{\Gamma}~~}}(\theta(t)).
\]
Bear in mind that $\mathcal{R}(\cdot)$ and $\mathcal{N}(\cdot)$ are invertible affine transformations. Note that the number of scheduling variables $n_\theta$ can either be considered as a prescribed value or chosen based on the accuracy index $\eta$. 

\section{Numerical illustration}
In this section, two numerical examples are provided to reveal the advantages of the proposed method in addition to comparison with some available approaches in the literature. 

\subsection{Example1}
\SA{This example is provided to evaluate the effectiveness of the proposed method for Problem 3 in this paper.} Consider an LPV system given by (\ref{StaSpa-Prob2}) with the following state-space matrices
\begin{multline}
\begin{pmat}[{|}]
\bar{\mathcal{A}}(\alpha) & \bar{\mathcal{B}}(\alpha) \cr\-
\bar{\mathcal{C}}(\alpha) & \bar{\mathcal{D}}(\alpha) \cr
\end{pmat}= 
\begin{pmat}[{.|}]
1+2\alpha_1  & 3+\alpha_2                      & 3\alpha_3+7\alpha_2 \cr
2+3\alpha_3 &20\alpha_1+5\alpha_2   & 1                                      \cr\-
\alpha_1       &  0                                        & 0                                     \cr
\end{pmat}
\end{multline}
with scheduling variable $\alpha(t)\in [0,2]\times [0,5] \times [-1,1]=\Omega_\alpha$.
The goal is to obtain an approximate LPV model (\ref{App-model}) with two scheduling variables $\theta\coloneqq\left[ \theta_1 ~\theta_2\right]^\top\in \Omega_\theta$  and with the minimum possible value of $\eta$, given by (\ref{Acc_Crit}), such that $\Omega_\theta$,  characterized by $\underline{\theta}_1$, $\overline{\theta}_1$, $\underline{\theta}_2$, and $\overline{\theta}_2$, has minimum-volume. 
\begin{figure}
	\begin{center}
		\includegraphics[width=9cm]{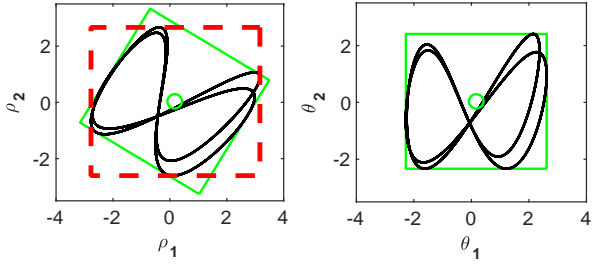}  
			\caption{Left plot: scheduling variables  $\rho_i$ (plotted over $\mathcal{D}_N$), $\Omega_\rho$ (red dashed box), and the minimum-volume enclosing box (green box). Right plot: scheduling variables  $\theta_i$ and $\Omega_\theta$. The figure demonstrates that computation of the minimal enclosing hyper rectangle and the proposed transformation results in smaller scheduling sets and hence reduced conservativeness of the resulting model.}
			\label{Fig_scheduling}
	\end{center}                                 
\end{figure}

Let us consider $T=10^{-3}$ and  scheduling trajectories 
\begin{align*}
\alpha_1(t)=2\sin^2(10t)&,  \quad
\alpha_2(t)=5\cos^2(20t+\frac{\pi}{5}),\\
\alpha_3(t)&=\sin(10t)\cos(20t).
\end{align*}
to generate $\mathcal{D}_N$ with $N=3000$. These trajectories adequately explore $\Omega_\alpha$ and represent the typical operation of the system we would like to preserve in our LPV model. Using the proposed method, an approximate model with two scheduling variables is obtained. In the left plot of Fig. \ref{Fig_scheduling}, the  scheduling variable vector $\rho$, $\Omega_\rho$, and the minimum-volume enclosing box is depicted.  Additionally, the scheduling variable vector $\theta$ and $\Omega_\theta$ is shown in the right plot  of Fig. \ref{Fig_scheduling}. The minimum-volume bounding box is obtained by the algorithm in \cite{Die20}. The centroid of the minimum-volume enclosing box around which the rotation is carried out is $(0.1688,0.0365)$. $\Omega_\theta$ can be characterized by $-2.2798\leq \theta_1\leq2.6174$ and $-2.3341\leq\theta_2\leq2.4071$. The volume of $\Omega_\rho$ and $\Omega_\theta$ are respectively $31.2870$ and $23.2186$. We have $\eta=54.4705$ for the proposed method. 

\begin{figure*}[t]
	\begin{center}
		\includegraphics[width=19cm]{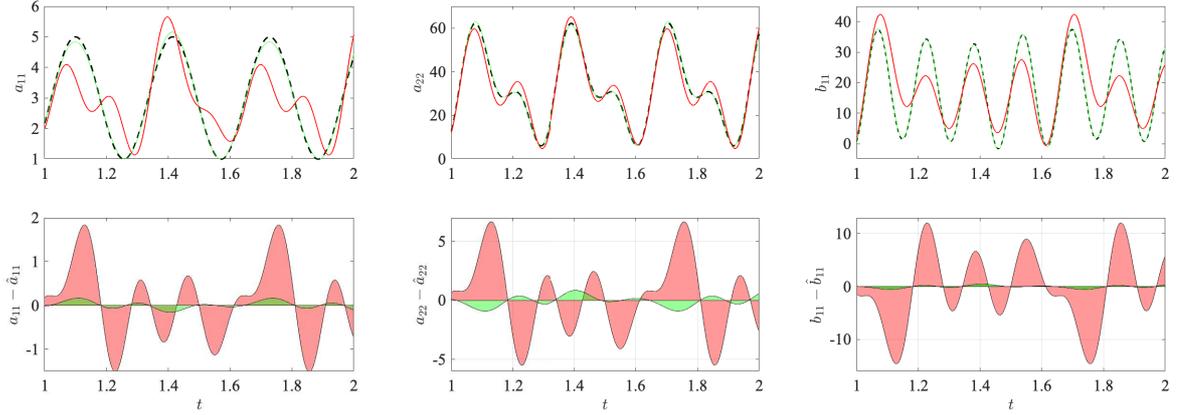}    
		\caption{Dashed lines: true trajectories of $a_{11}$, $a_{22}$, and $b_{11}$. Thick lines (red): approximate trajectories obtained by the method of \cite{KwiWer08}. Thin lines (green): approximate trajectories obtained by the proposed method. The figures show that the proposed method achieves significantly lower approximation error of the system with 2 scheduling variables than the method in \cite{KwiWer08}.}
		\label{Fig_Comparison}
	\end{center}
	\hrulefill
\end{figure*}

For comparison purposes, the proposed method in \cite{KwiWer08} is also applied on this system to obtain an approximate model with two scheduling variables. To visualize the results, variation of some elements of $\mathcal{A}(\alpha(t))$: $a_{11}=1+2\alpha_1$, $a_{22}=20\alpha_1+5\alpha_2 $, and $b_{11}=3\alpha_3+7\alpha_2$ and their approximate counterparts are shown in Fig. \ref{Fig_Comparison} for a time interval  between $t=1$ and $t=2$. Obviously, the proposed method provides a better approximation for the original LPV system with the same number of the scheduling variables.  It is worth to mention that for the method in \cite{KwiWer08}, the obtained value for $\eta$, given by (\ref{Acc_Crit}), equals to $\eta=68.2811$, which is greater than that of the proposed method ($\eta=54.4705$). This also reinforces the superiority of the proposed method. 

\subsection{Example 2} \label{Example2}

\SA{In this example, the applicability of the proposed method for Problem 1 is numerically investigated.} Consider a nonlinear system given by (\ref{StaSpa_rep})  as follows:
\begin{multline} \label{Init_Model}
\left[ \begin{array} {cc}
\mathcal{A}(x,u) & \mathcal{B}(x,u) \\
\mathcal{C}(x,u) & \mathcal{D}(x,u)
\end{array}
\right]= 
\begin{pmat}[{.|}]
2\sin(x_1)+1   & 3x_1+5       & 0                  \cr
x_1                &0                     &  1  \cr\-
\sin(x_1)                       &  2x_1           & 0                  \cr
\end{pmat}
\end{multline}
It is assumed that $-\frac{\pi}{2}\leq x_1(t)\leq\frac{\pi}{2}$. In the proposed method, after constructing normalized matrix $\bar{\Pi}_\alpha$ using the sampling period $T=0.01$, one can see that the singular values are 
\[
\sigma_1=39.5533,\quad\sigma_2=2.3526,\quad\sigma_3=\sigma_4=\sigma_5=0,
\]
which implies that using just two scheduling variables an equivalent LPV embedding of the system is available. Obviously, it is expected since all the elements of $\mathcal{E}(x(t),u(t))$ are affine functions of the terms $\sin(x_1(t))$ and $x_1(t)$ which can be considered as the scheduling variables. Note that the zero terms in  $\mathcal{E}(x(t),u(t))$ are excluded from construction of $\Gamma(\alpha(t))$ without any loss of generality.  Applying the proposed method and considering two scheduling variables yields an exact LPV model (\ref{App-model}) with $\hat{L}(\theta(t))$ as
\begin{equation} \label{L_hat_Example2}
\hspace{-1mm}\begin{pmat}[{.|}]
0.6337\theta_1+0.7773\theta_2+1   & 1.2226\theta_1-0.9968\theta_2+5       & 0                  \cr
0.4075\theta_1-0.3323\theta_2        &0                                                         & 1    \cr\-
0.3169\theta_1+0.3887\theta_2      &  0.8151\theta_1-0.6645\theta_2                                       & 0                  \cr
\end{pmat}
\end{equation}
where the scheduling variables $\theta_1$ and $\theta_2$  \SA{ are defined by the following affine mapping of the elements of (\ref{Init_Model}): } 
\SA{
\begin{align*}
&\left[\begin{array} {c} \theta_1 \\ \theta_2 \end{array}\right]= \\ &\left[\begin{array}{cc} 
0.3150  & 0.3864\\
0.1638  & -0.1335   \\
0.4913 & -0.4006  \\
0.6301  & 0.7728\\
0.2457 &  -0.2003
 \end{array} \right]^\top\left[\begin{array}{c} 2\sin(x_1)+1 \\ 3x_1+5 \\x_1 \\\sin(x_1) \\ 2x_1 \end{array} \right]+ 
 \left[\begin{array}{c} -1.1339 \\
 0.2812 \end{array}\right]
\end{align*}
The above given construction implies the scheduling map
}
\[
\mu(x(t))=\left\{ 
\begin{aligned}
\theta_1(t)&=1.2601\sin(x_1(t))+1.4740x_1(t), \\
\theta_2(t)&=1.5456\sin(x_1(t))-1.2017x_1(t) .
\end{aligned}
\right.
\]
If $\theta_1(t)$ and $\theta_2(t)$ are substituted with the scheduling map $\mu$ in $\hat{L}(\theta(t))$, then the same nonlinear model (\ref{Init_Model}) is obtained. However, note that the proposed method is an automated affine LPV embedding approach for the  nonlinear systems. 

One can see that $\sigma_2$ is negligible in comparison with $\sigma_1$. Therefore, it is also possible to obtain an approximately accurate LPV model with just one scheduling variable for this system. After applying the proposed method, an affine LPV model is obtained:
\begin{equation}
\hat{L}(\theta)= 
\begin{pmat}[{.|}]
0.6337\theta_1+1            & 1.2226\theta_1+5       & 0                           \cr
0.4075\theta_1               & 0                            &  1     \cr\-
0.3169\theta_1               &  0.8151\theta_1      & 0                            \cr
\end{pmat}
\end{equation}
with $\theta_1(t)=\mu(x(t))=1.2601\sin(x_1(t))+1.4740x_1(t)$.

\begin{figure}
	\begin{center}
		\includegraphics[width=9cm]{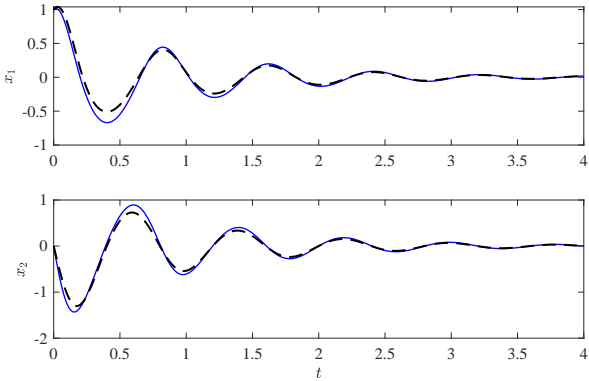}    
		\caption{The state trajectories of the LPV model (blue lines) and the nonlinear system (dashed black lines) in a  time-domain simulation using a gain-scheduled state feedback controller for Example \ref{Example2}. The responses show that the LPV model with a single scheduling variable has highly similar closed-loop response as the nonlinear system. }
		\label{Exa2_Sim}                                 
	\end{center}                                 
\end{figure}
As these systems are unstable, for comparison purposes, the response of the nonlinear and the affine LPV system (with one scheduling variable) are computed under the same gain-scheduled state feedback controller. Fig. \ref{Exa2_Sim} reveals good coincidence between the time-domain simulation results starting from the initial state $x(0)=[1~0]^\top$. 
\section{Experimental example}\label{sec:exp}
\begin{figure}
	\begin{center}
		\includegraphics[width=3cm]{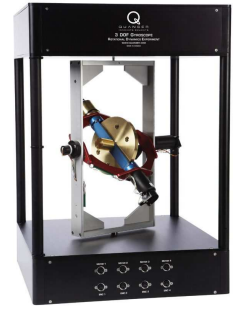}    
		\caption{The gyroscope experimental setup by Quanser}
		\label{Gyro_Fig}                                 
	\end{center}                                 
\end{figure}

In this section,  the proposed method is applied on a 3-DOF gyroscope system by Quanser, shown in Fig. \ref{Gyro_Fig}. Using a first-principle model of the system and measured data, an LPV model of the system is constructed with our method. Subsequently, exploiting the obtained model, a full-order gain-scheduled output feedback controller is designed and applied on the setup. Converting the motion model of the gyroscope to an LPV form is challenging and results in an excessive number of scheduling variables \cite{HofWer15Conf2}, so  obtaining an LPV model with low number of scheduling variables is an achievement in itself by the proposed method. 

\subsection{Plant description}

The gyroscope consists of a golden flywheel mounted inside an inner blue gimbal which in turn is mounted inside an outer red gimbal. The red gimbal is attached to a rotating silver frame. In the experiment considered in this paper, it is supposed that the rectangular silver frame is fixed. The blue and red gimbals can be actuated about their rotation axes using DC motors and the angular position of both gimbals are measured using optical encoders. The flywheel is actuated using another motor. 

In our experiment, the flywheel is regulated by a controller which follows an unknown reference signal. The objective is to achieve servo control of the red and blue frames together by rejecting the disturbance generated by the change of the velocity of the flywheel. Let $q_2(t)$ and $q_3(t)$ represent the angular position of the blue and red gimbals, respectively. Moreover, the angular velocity of the flywheel, the blue gimbal, and the red gimbal are denoted by $\dot{q}_1(t)$, $\dot{q}_2(t)$, and $\dot{q}_3(t)$. The torque applied on the blue and red gimbals are given by $\tau_b(t)$ and $\tau_r(t)$. 

\subsection{LPV modeling} \label{LPV_Model_sec}
\begin{figure}
	\begin{center}
		\includegraphics[width=8cm]{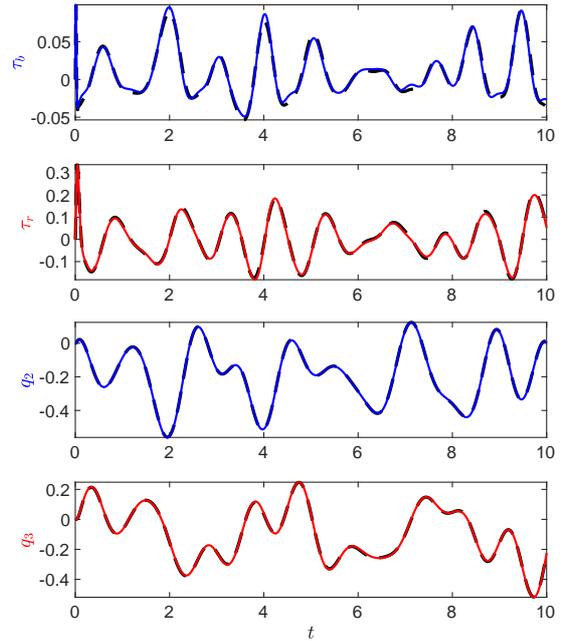}    
		\caption{Validation plots of input torques $\tau_b$, $\tau_r$ and angular positions $q_2$, $q_3$  of the original nonlinear system (solid-red lines for gimbal red and solid-blue lines for gimbal blue)  and the LPV model (dashed-black line).}
		\label{ModelValidationFig}                                 
	\end{center}                                 
\end{figure}

\begin{figure}
	\begin{center}
		\includegraphics[width=9cm]{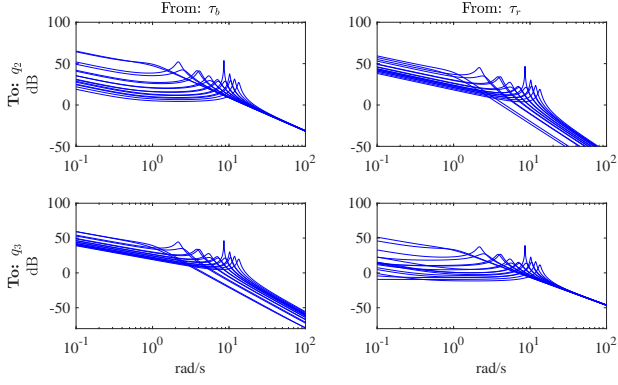}
		\caption{Magnitude plots of the open-loop frequency responses of the LPV model for frozen (constant) scheduling \SA{variables}, where significant gain and pole variations can be observed over the scheduling range.}
		\label{OpenLoopFig}  
	\end{center}
	\end{figure}

Through the Euler-Lagrange equations and frames defined for the rational bodies in the gyroscope, a dynamical motion model of the system is derived:
\begin{equation} \label{gyro_or}
\mathcal{M}(q)\ddot{q} + \mathcal{C}(q,\dot{q})\dot{q} = \tau.
\end{equation}
where $\mathcal{M}(q)$ is the inertia matrix and $C(q,\dot{q})$ is the Coriolis matrix whose elements are derived as the result of the summation of the Christoffel symbols and the generalized angular velocities. The corresponding coefficients of these matrix functions are identified using measured data form the physical system and prediction error minimization with nonlinear optimization. As the flywheel can be seen as a separate subsystem, an NL-SS model of \eqref{gyro_or} w.r.t. the dynamics of the blue and red  gimbals is derived in terms of \eqref{eq:fact} with
\begin{equation} \label{Non_Gyro}
\left[ \begin{array} {cc}\!
\mathcal{A}(x,q_1)\! &\! \mathcal{B}(x)\! \\
\mathcal{C} & \mathcal{D}
\end{array}
\right]\!\!=\! \!
\begin{pmat}[{...|.}]
0 & 0 & 1 & 0 &0 & 0 \cr 0 & 0 & 0 & 1 & 0  & 0 \cr 0 & 0 & f_1(x,q_1) & f_2(x,q_1) &g_1(x) & g_2(x) \cr 0 & 0 & f_3(x,q_1) & f_4(x,q_1)  & g_3(x) & g_4(x) \cr\-
1 & 0& 0& 0  & 0 &0 \cr
0 & 1 & 0& 0 & 0&0 \cr
\end{pmat}
\end{equation}
with $x \coloneqq[\begin{array}{cccc} q_2 & q_3 &  \dot{q}_2 & \dot{q}_3\end{array} ]^\top$ and $u=[\begin{array}{cc} \tau_b & \tau_r\end{array}]^\top$. Due to the division by $\mathcal{M}(q)$ to derive \eqref{Non_Gyro}, the rather complicated rational trigonometrical expressions of $f_1,\ldots,f_4,$ which all depend on $(q_2,q_3,\dot{q}_1,\dot{q}_2,\dot{q}_3),$ and also $g_1,\ldots,g_4,$ which depend on $(q_2,q_3),$ are not given here. The interested reader can obtain them from the equations provided in \cite{BloTot19}. (\ref{Non_Gyro}) is a difficult nonlinear model for which we would like to obtain a low complexity affine LPV model (Problem 1). Using the proposed method with $n_\theta=2$ an affine LPV model is obtained:
\begin{align}
\dot{x}_p(t)&=A_p(\theta(t))x_p(t)+B_p(\theta(t))u(t), \\
y(t)&=C_px_p(t),
\end{align}
where $\theta\coloneqq \left[\begin{array}{cc} \theta_1 & \theta_2 \end{array}\right]^\top$. The corresponding matrices in the affine $A_p(\theta)$,  $B_p(\theta)$, and the constant matrix $C_p$ are not included here  for the sake of brevity.  Moreover, the following bounds on the scheduling variables are obtained: 
\begin{equation*}
-4.5577\leq\theta_1(t)\leq2.3802,\quad~ -3.4307\leq\theta_2(t)\leq3.4885
\end{equation*}
Additionally, one can readily compute the  bounds on the derivative of the scheduling variables given below which are required for the controller synthesis in the next sections. 
\begin{equation*}
-70.9504\leq\dot{\theta}_1(t)\leq 81.4165, \quad -16.9968\leq \dot{\theta}_2(t)\leq 33.9214
\end{equation*}
Thus,  $\dot{\theta}(t)$ also lies  in a hyper rectangle 
$
\dot{\theta}(t)\in \Lambda_\theta.
$
The quality of the obtained LPV model is assessed in a closed-loop simulation study with an LPV controller designed in Section \ref{Cont_synthesis_sec} and compared with the closed loop response of the nonlinear system operated with the same controller. In case of the nonlinear gyroscope model, 
the angular velocity of the disc $\dot{q}_1$ is regulated by a second controller to track a sinusoidal reference. Two multisine signals, different from those employed to obtain $\mathcal{D}_N$ and construct $\Pi_\alpha$, are used as the desired reference signals for $q_2$ and $q_3$ in the closed-loop simulation. The related results for $\tau_b$, $\tau_r$, $q_2$, and $q_3$ are shown in Fig. \ref{ModelValidationFig}. The \emph{root mean square error} (RMSE) of the obtained response of the LPV model w.r.t. the true system response is given in Table \ref{RMSE_table}.  Obviously, the LPV model well captures the dynamics of the original system. Moreover, magnitude plots of the frequency response of the LPV model are depicted in  Fig. \ref{OpenLoopFig} for some frozen scheduling variables in the related intervals, which clearly reveals the significant variation of the model over the scheduling set. 
\begin{figure}
	\begin{center}
		\includegraphics[width=8cm]{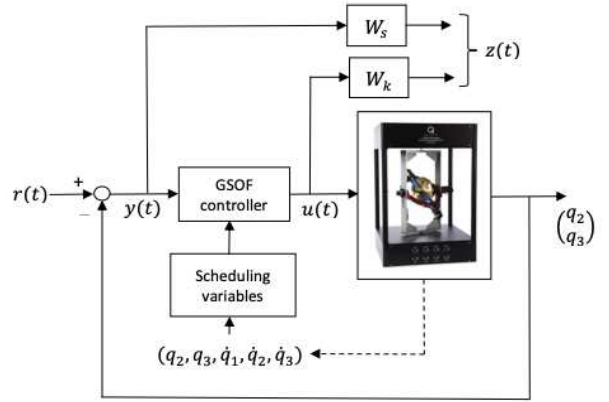}    
		\caption{Closed-loop control setup for the gyroscope.}
		\label{ClosLoopSetup}                                 
	\end{center}                                 
\end{figure}
\begin{figure}
	\begin{center}
		\includegraphics[width=9cm]{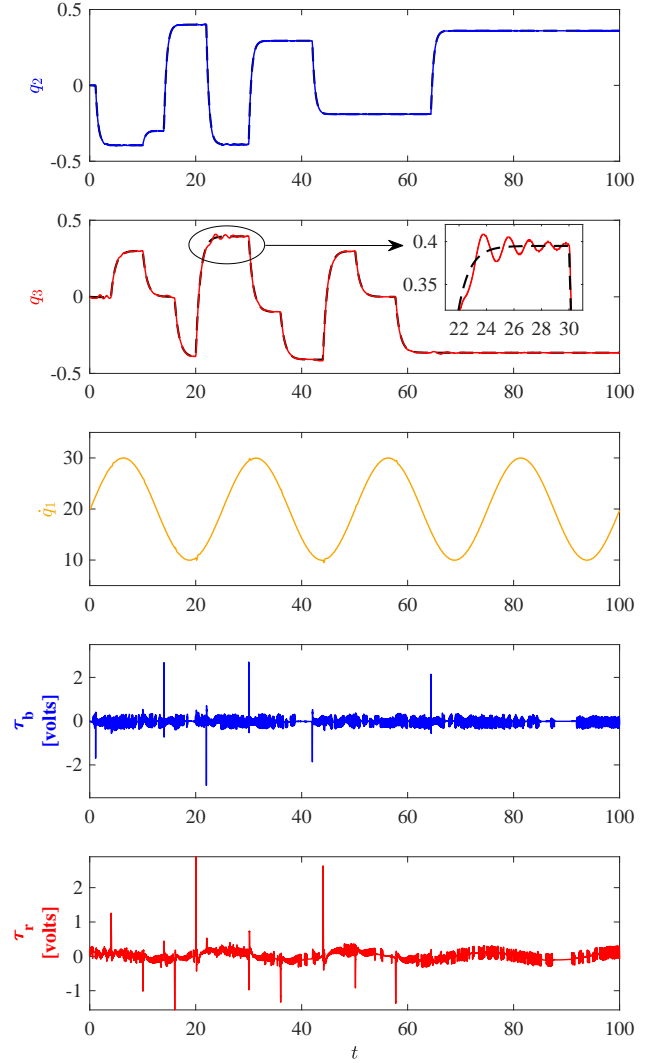}    
		\caption{Experimental results of the designed LPV controller with the gyroscope. Dashed lines: the desired reference. Solid lines: the obtained results from the real setup. The LPV controller designed based on the low complexity LPV embedding of the plant using the proposed approach of the paper shows adequate tracking performance with this highly nonlinear system.}
		\label{ExperResultFig}                                 
	\end{center}                                 
\end{figure}
\begin{table}
	\centering
	\caption{RMSE values of the obtained closed-loop simulation response of the LPV model w.r.t. the nonlinear gyroscope model operated by the same controller.}
	\begin{tabular}{ccccc}
		\hline
	           &  $\tau_2$           &  $\tau_3$           & $q_2$              &  $q_3$    \\
		\hline
		       RMSE              &  0.0036             &  0.0049              & 0.0013             &  0.0029 \\
		\hline           
		\label{RMSE_table}
	\end{tabular}
\end{table}

\subsection{Control synthesis} \label{Cont_synthesis_sec}
In this section, using the obtained LPV model in Section \ref{LPV_Model_sec}, a gain-scheduled full-order output feedback controller is sought for the gyroscope based on the closed-loop setup shown in Fig. \ref{ClosLoopSetup}. Roughly speaking, the $H_\infty$-type performance, more precisely the induced $\mathcal{L}_2$-gain performance, is considered to shape the frozen sensitivity and control sensitivity functions of the feedback system. To obtain good tracking performance, while the actuator constraints are taken into account, the following weighting functions are employed: 
\begin{align*}
W_s&=\left[\begin{array}{cc} \frac{32s+20000}{1.6s+1} & 0 \\ 0  &  \frac{11.2s+7000}{1.6s+1} \end{array}\right],\\
W_k&=\left[\begin{array}{cc} \frac{0.024s+1.5}{0.0016s+1} & 0 \\0  & \frac{0.096s+6}{0.0016s+1} \end{array}\right],
\end{align*}
\SA{The weighting function $W_s$  has been designed based on the standard mixed-sensitivity shaping approach to shape the frozen sensitivity frequency responses of the closed-loop system. $W_s$ is chosen such that corresponds to a low-pass filter with bandwidth 0.7 rad/s  and low frequency gain of about 80 dB to ensure good disturbance attenuation, fast response time and less than 20\% of expected overshoot.  Similarly the control sensitivity weighting function, $W_k$, has been chosen considering the amplitude and frequency constraints on the voltage to be applied to the DC motors of the experimental setup.}
The open-loop weighted plant displayed  in Fig. \ref{ClosLoopSetup} (by removing the controller) can be expressed as: 
\begin{align} \nonumber \label{AugmentSys}
\dot{x}(t)&=A(\theta(t))x(t)+B_r(\theta(t))r(t)+B_u(\theta(t))u(t), \\
z(t)&=C_z(\theta(t))x(t)+D_r(\theta(t))r(t)+D_u(\theta(t))u(t), \\ \nonumber
y(t)&=C_y(\theta(t))x(t)+D_y(\theta(t))r(t),
\end{align}
where $x(t) \in\Real^n$, $r(t)\in \Real^m$, $u(t) \in \Real^p$, $z(t)\in \Real^q$, and $y(t)\in \Real^r$ respectively denote the state vector, the external  input, the control input, the performance output, and the measured output of the system, with 
\begin{multline}
\left[
\begin{array}{ccc}
A(\theta(t)) & B_r(\theta(t)) & B_u(\theta(t)) \\
C_z(\theta(t)) & D_r(\theta(t)) & D_u(\theta(t)) \\
C_y(\theta(t)) & D_y(\theta(t)) &0
\end{array}
\right]= \\
\begin{pmat}[{..||}]
A_p(\theta(t))        & 0      & 0     &    0    &     B_p(\theta(t)) \cr
-B_sC_p(\theta(t)) & A_s  & 0     &    B_s &     0                    \cr
0                           & 0     & A_k  &    0    &     B_k                 \cr\-
-D_sC_p(\theta(t)) & C_s &   0    &   D_s  &    0                    \cr
0                           & 0    &   C_k &  0       &  D_k                  \cr\-
-C_p(\theta(t))      & 0     &  0     &  I        &        0                 \cr
\end{pmat}
\end{multline}
where $(A_s,B_s,C_s,D_s)$ and $(A_k,B_k,C_k,D_k)$ are the state-space realizations of $W_s$ and $W_k$, respectively. 

The goal is to design the full-order gain-scheduled controller
\begin{equation}
\mathcal{K}(\theta):~
\left\{\begin{aligned} \label{Controlle_SS} \nonumber
\dot{x}_c(t)&=A_c(\theta(t))x_c(t)+B_c(\theta(t))y(t) \\ 
u(t)&= C_c(\theta(t))x_c(t)+D_c(\theta(t))y(t)
\end{aligned}
\right.
\end{equation}
such that it stabilizes the closed-loop system and assures an upper bound $\gamma$ on the induced $\mathcal{L}_2$-gain performance. 
To design the controller, we use the method detailed in \cite{SadIJRNC18}.

The  resulting parameter-dependent LMI problems are solved through finite-dimensional LMI relaxation exploiting homogeneous polynomial matrices inspired by the method of \cite{OliPer07}. To this end, YALMIP \cite{Lof04} and ROLMIP \cite{ROLMIP} interfaces for the LMI solver MOSEK \cite{mosek} are employed. With a $\lambda=0.001$ an 8th-order controller is obtained with a guaranteed $\mathcal{L}_2$-gain performance bound of $240.81$. 
\subsection{Experimental results} 
In this section, the gain-scheduled controller designed based on the LPV model obtained via the proposed method in this paper  is experimentally validated on the laboratory setup. The controller is described in block diagrams in MATLAB/Simulink. Then using a dSPACE board that implements a real time interface, it is applied on the setup. The angular velocity of the flywheel is made to track a sinusoidal reference between $10$ to $30$ rad/sec.  Since the movements of blue and red gimbals do not considerably affect the rotational velocity  of the flywheel, a simple proportional controller is employed to ensure the tracking. The change of the velocity of the flywheel is considered as the exogenous disturbance, and have to be rejected by the controller. The reference trajectories for gimbals blue and red, i.e. $q_2$ and $q_3$, change within $\pm0.4$ rad and are designed to cover different positions of blue and red gimbals with respect to each other. The desired and actual positions of the gimbals, the angular velocity of the flywheel, and the control input signals are shown in Fig. \ref{ExperResultFig}. One can see that both gimbals track successfully the related reference trajectories while the input torques $\tau_b$ and $\tau_r$ remain within acceptable levels.  

\section{Conclusion}
A novel method taking advantage of PCA is devised in this paper for LPV embedding of nonlinear systems. Contrary to the available methods for LPV embedding of nonlinear models based on PCA, the PCA is applied on a data matrix consisting of the trajectories of all elements of the state-space matrices not solely the scheduling variables. Furthermore, by finding the minimum bounding box of the scheduling variables and its proper rotation, the conservativeness related to  over bounding the admissible region of the scheduling variables is reduced  by introducing  an invertible transformation to find   a new set of scheduling variables. In addition to academic examples, the proposed method is deployed to generate an LPV model for a 3-DOF gyroscope. Assessment of the LPV model  and evaluation of the closed-loop performance, which is obtained by a designed gain-scheduled controller exploiting the developed LPV model, successfully demonstrate the applicability of the developed method in practice. 

\section*{Acknowledgment}
This work has received funding from the European Research Council (ERC) under the European Union’s Horizon 2020 research and innovation programme (grant agreement nr. 714663).

\bibliographystyle{ifacconf}
\bibliography{MyCollection1}             
\end{document}